\documentclass[aps,prl,preprint,showpacs,groupedaddress,floatfix]{revtex4}
 
\usepackage[dvips]{graphicx}
\usepackage{amssymb}
 
\begin{document}

\title{Interplay between pairing and exchange in small metallic dots}

\author{G.~Falci$^{(1)}$, Rosario Fazio$^{(2,3)}$, and A. Mastellone$^{(1,3)}$}
\affiliation{
	$^{(1)}$Dipartimento di Metodologie Fisiche e Chimiche
	(DMFCI), INFM, UdR di Catania,
	Universit\`a di Catania, viale A. Doria~6, I-95125 Catania, Italy\\
	$^{(2)}$Scuola Normale Superiore, INFM, UdR di Pisa, 
        Piazza dei Cavalieri 7, I-56126 Pisa, Italy\\
	$^{(3)}$NEST-INFM, 
        National Enterprise for nanoScience and nanoTechnology, Italy}

\date{\today}

\begin{abstract}
We study the effects of the mesoscopic fluctuations on the competition
between exchange and pairing interactions in ultrasmall metallic dots 
when the mean level spacing $\delta$ is comparable or larger than the 
BCS pairing energy $\Delta$. Due to mesoscopic fluctuations, the probability 
to have a non-zero spin ground state may be non-vanishing and shows universal 
features related to both level statistics and interaction. Sample to 
sample fluctuations of the renormalized pairing are enlightened.
\end{abstract}

\maketitle

The competition of superconductivity and ferromagnetism and their 
possible coexistence in bulk materials has been intensively investigated 
in the past and it lead to a very rich phase diagram~\cite{ferrosuper}. 
In addition to the suppression of superconductivity due to time reversal 
symmetric breaking, new phases with non-uniform magnetic and 
superconducting ordering were predicted~\cite{fsreview}.
This picture does not hold when the size of the sample is reduced such 
that the average level spacing $\delta$ becomes comparable with the 
energy scales related to the onset of macroscopic order.
In this case fluctuation (both thermal and quantum) effects become 
very important and even the definition of ordered phase is elusive.
For instance a small superconductor will not posses a 
fully developed gap~\cite{pairing} and the spin of a ferromagnet will not be 
macroscopically large~\cite{exchange}. Nevertheless distinct features 
reminiscent of the macroscopic order can be traced even in nanosized grains. 

In this Letter we investigate the competition between 
exchange and pairing in small metallic grains. Our approach starts 
from the results of a recent work of Kurland 
{\em et al.}~\cite{kualal} who have shown that universal features of 
electrons in isolated mesoscopic grains are accounted for by the 
Hamiltonian (see also Refs.~\cite{blanter96,aleiner97}) 
\begin{equation}
{\cal H} = \sum_{\alpha=1}^{\Omega} \sum_{\sigma=\{\uparrow,\downarrow\}}
\epsilon_{\alpha} c_{\alpha,\sigma}^{\dagger} c_{\alpha,\sigma}
+ E_C \hat{N}^2 - \lambda \hat{T}^{\dagger} \hat{T} - J \hat{S}^2 \; .
\label{hamiltonian}
\end{equation}
This description is valid in the limit of very large
dimensionless conductance $g=E_T/\delta$, being $E_T$ the Thouless energy and
$\delta$ the mean level spacing. In Eq.(\ref{hamiltonian}) the first term is 
the kinetic energy while the electron-electron interaction is expressed 
as a sum of three terms, which describe respectively charging, pairing 
and exchange interactions. 
The index $\alpha$ spans a shell of $\Omega$ doubly degenerate ($\sigma=\pm$) 
time reversed  single particle states of energy 
$\epsilon_\alpha$, and $c_{\alpha,\sigma}$ ($c^{\dagger}_{\alpha,\sigma}$ )
are the corresponding annihilation (creation) operators.
The energies $\epsilon_\alpha$ are distributed according to the Gaussian
Orthogonal Ensemble (GOE), which describes the case of time-reversal
symmetry with no spin-orbit coupling~\cite{mehta}.
The interaction depends only on the collective variables
$N=\sum_{\sigma} 
%=\{\uparrow,\downarrow\}} 
c_{\alpha,\sigma}^{\dagger}
c_{\alpha,\sigma}$ (the number operator),
$T = \sum_{\alpha=1}^{\Omega}c_{\alpha,-}c_{\alpha,+}$ (pair creation 
operator),
and $\vec{S} = \sum_{\alpha=1}^{\Omega}c_{\alpha,\sigma}^{\dagger}
\vec{\sigma}_{\sigma,\sigma'} c_{\alpha,\sigma'}$ (the total spin operator,
$\sigma$ are the Pauli matrices).
For isolated grains $N$ is fixed and the charging term can be ignored. 
Pairing interaction tends to favour the formation of
spin singlets, while exchange tends to favour maximal spin, so they
compete in determining the spin ordering.

A clear picture is already emerging  in the case where either pairing 
or exchange is present. In the absence of exchange interactions ($J=0$),
the Hamiltonian in Eq.(\ref{hamiltonian}) describes pairing 
correlations in small metallic grains~\cite{pairing}. Although no phase 
transition occurs, signatures of pairing correlations have been 
detected~\cite{BRT} even in nanosized grains.
For this model it has been shown~\cite{matlar,mastellone} that the 
low-energy properties are universal functions of the ratio
$\delta/\Delta =2 \sinh(\delta/\lambda)/\Omega$
($\Delta$ is the BCS gap value). Upon increasing the size of 
the grains, hence decreasing the ratio $\delta/\Delta$, there is a  
crossover~\cite{mastellone} between the case of ultrasmall grains 
($\delta \gg \Delta$) where pairing produces strong quantum 
fluctuations~\cite{matlar}, and a regime ($\delta \ll \Delta$), where 
the BCS mean-field description remains valid~\cite{vondelft}. 
Signatures of pairing correlations may be detectable in thermodynamic 
quantities~\cite{thermo}, even for ultrasmall grains.
In absence of paring ($\lambda=0$) the properties of the model of 
Eq.(\ref{hamiltonian}) are determined by the interplay between the
kinetic term, which favours Pauli filling of the levels and zero total 
spin in ground state, and  exchange one which tends to maximize $S$ and 
eventually leads to the Stoner instability for $J \ge \delta$. 
However in mesoscopic samples~\cite{kualal,baranger,brouwer,jasto} 
it is possible
to find individual grains with a cluster of $2S$ closely spaced levels
around the Fermi energy, whose ground state may have spin $S$ even for 
$J \ll \delta$. The probability $P_S(J/\delta)$ 
of spin-$S$ ground state directly 
reflects the universal properties of the level statistics~\cite{folk}.

In order to describe the interplay between superconductivity and 
ferromagnetism in small grains we study how the tendency to magnetic 
ordering is reduced when pairing is gradually increased. 
Hence, in the same spirit as in Ref.~\cite{folk}, we consider
the probability $P_S$ of finding a spin-$S$ ground state in the regime 
$J, \lambda \ll \delta$, which is in turn related to 
the spontaneous magnetization of an ensemble of grains.
All the results that will be presented are obtained in
the half-filling scheme, where the numbers of electrons $N$ is equal to 
the number of levels $\Omega$.
For an ensemble of normal grains $P_S(J/\delta)$ 
for $J \ll \delta$ is non-zero
due to grains  with a cluster of $2S$ close levels around the 
Fermi energy~\cite{folk}. The same picture applies 
when pairing interaction is present except that, since the energy 
balance between spin $S$ states involves the pairing correlation energy, 
one should take into account contributions coming from the entire shell 
of $\Omega$ levels.
In the weak coupling limit $J,\lambda \ll \delta$ there is a simple way to
circumvent this problem, namely to consider a shell of $2S$ levels with a 
renormalized pairing coupling $\tilde{\lambda}_{2S}$ (see Ref.~\cite{berhal}).
Notice that whereas no sample to sample fluctuations affect the bare 
$\lambda$, the actual level distribution may produce fluctuations in 
$\tilde{\lambda}_{2S}$, which we ignore at the moment. With this hypothesis 
the probability to have spin-$S$ in the ground state, for
$0 < J-J^*_S \ll \lambda \ll \delta$, is
\begin{equation}
P_S \left ( \frac{J}{\delta} \right ) = C_S \left ( \frac{J}{\delta} 
\right )^{\alpha_S} 
\; \left ( \frac{J- J^*_S}{\delta} \right ) ^{\alpha_S},
\label{p_s}
\end{equation}
where $\alpha_S=(S+1)(2S-1)/2$, 
$J^*_S$ is the threshold value of the exchange below which the spin 
probability vanishes and depends on the spin and the pairing
($J^*_1=\tilde{\lambda}_2$, $J^*_{3/2}=2\tilde{\lambda}_3/3$) and the
$C_S$ are dimensionless constant depending only on the spin 
$S$ ($C_1 = \pi^2/3$, $C_{3/2} = 9 \pi^4 /50$) which directly reflect the
universal statistical properties of the GOE level distribution.
Compared to the case where pairing is absent~\cite{kualal}, the exponent 
$\alpha_S$ is halved.

\begin{figure}
\includegraphics[width=100mm]{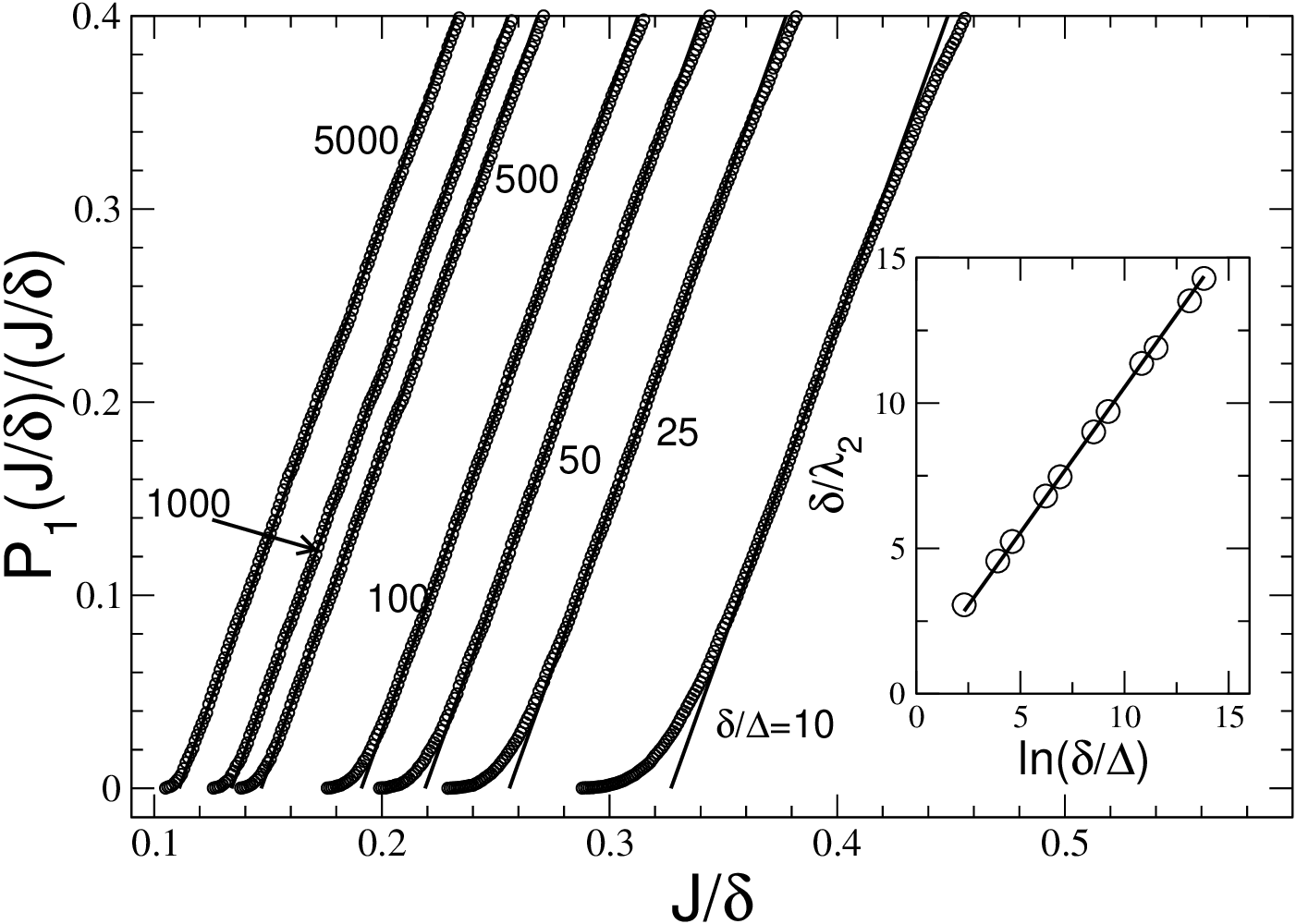}
\caption{\label{peven} The $P_1(J/\delta)$ probabilities in the
case $\Omega=30$ for the different values of the ratio
$\delta/\Delta$ (circles). The lines represent the fits to the 
data using the expression given in Eq.(\ref{p_s}). 
In the inset, the numerical data of the renormalized 
pairing constant (circles) extracted from $J_1^{\star}$ are plotted 
as a function of $\delta/\Delta$. 
They coincide with the value of the renormalized pairing 
$\tilde{\lambda}_2=\delta/\ln(a_1 \delta/\Delta)$ (with $a_1=1.721$).}
\end{figure}

\begin{figure}
\includegraphics[width=100mm]{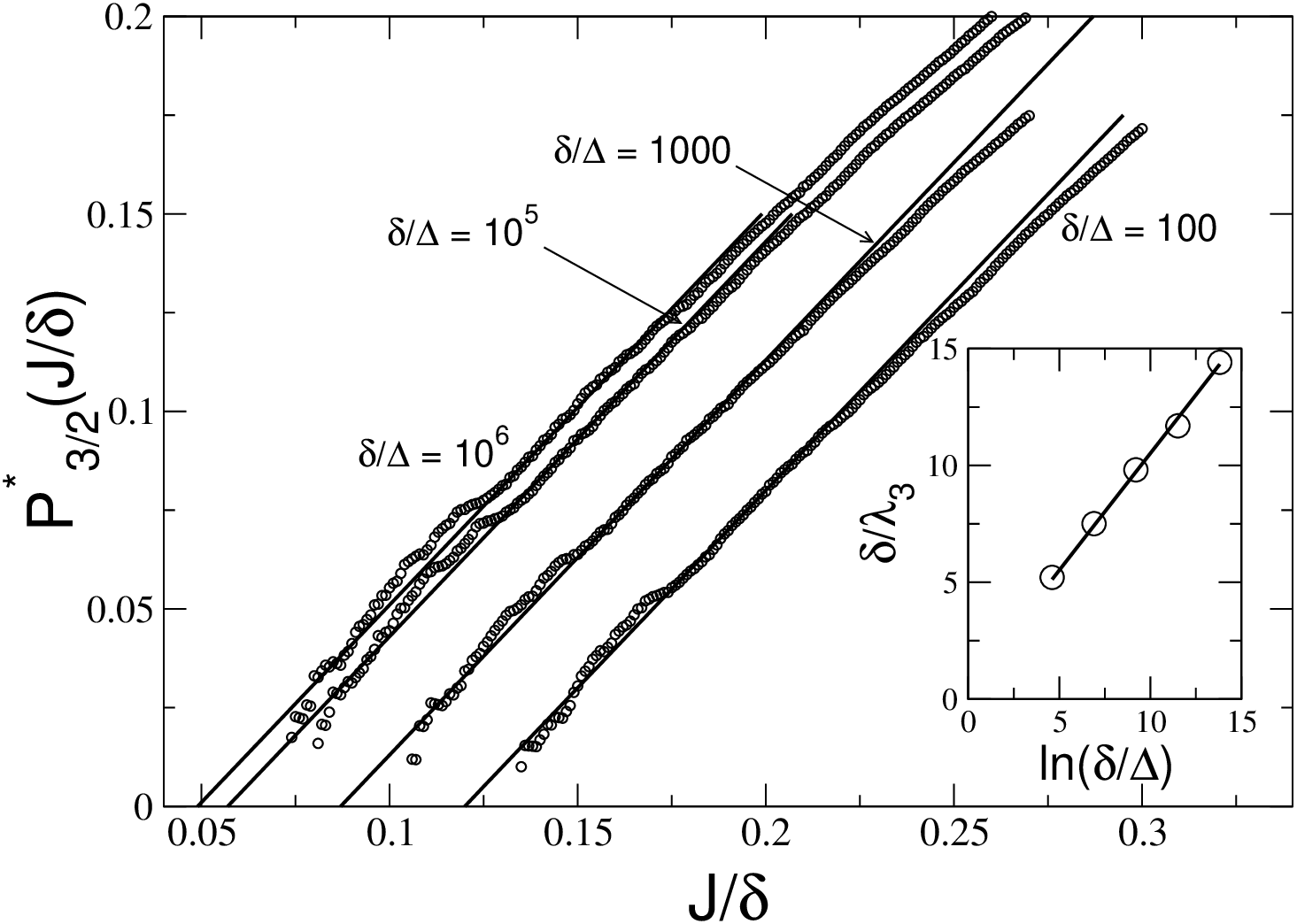}
\caption{\label{podd} The probabilities in the case $\Omega=31$
for the different values of the ratio $\delta/\Delta$. 
For convenience we plotted
$P^*_{3/2}(J/\delta)=[1350 P_{3/2}(J/\delta)/\pi^4]^{(2/5)}/(9J/\delta)
=(J-2\tilde{\lambda}_3/3)/\delta$.
The circles represent the numerical data, and the lines
the linear fits. In the inset, the numerical data threshold value as a 
function of the ratio $\delta/\Delta$. The scaling behaviour is 
obtained as in the previous figure with $a_{3/2}=1.679$.}
\end{figure}

We now check these results with numerical calculations. 
We have considered grains with size up to 
$\Omega=30,31$, which are large enough to show the universal behavior of the
pairing interaction~\cite{mastellone}, as in grains with much larger $\Omega$.
We considered ensembles of grains whose single particle spectra 
$\{\epsilon_{\alpha},\alpha=1,\dots,\Omega\}$ realize 
the GOE level statistics. To this end 
we diagonalized a set of $5\Omega$x$5\Omega$ real
orthogonal random matrix taking out the central $\Omega$ eigenvalues, 
in order to avoid edge effects. 
Then we found the many-particle energies of Eq.(\ref{hamiltonian}) 
by using the Richardson exact solution~\cite{richardson} for 
larger systems ($\Omega \ge 20$), whereas for smaller systems we used the 
standard numerical diagonalization of the Hamiltonian or the Lanczos method.
For systems with even (odd) $N$ we studied the probability $P_1$ ($P_{3/2}$) 
of a ground state with spin 1 (3/2) using a set of $10^5$ ($10^6$)  
level configurations (systems with odd $N$ require more statistics 
because $P_{3/2}$ is smaller).
Results are shown in Fig.\ref{peven} and  Fig.\ref{podd} for values of 
$J/\delta$ such that the probability that the ground state has larger spin 
($S>1$ or $S>3/2$) is negligible. 
The result given in Eq.(\ref{p_s}) reproduces quite well the numerical 
data, except for smaller $\delta/\Delta$ (large grains) and near $J_S^*$, 
a regime which we discuss later. 
Thus we conclude that $P_1$ and $P_{3/2}$ show the 
universal features of the level statistics predicted in Eq.(\ref{p_s}), 
namely the coefficient $C_S$ and the power law as a function of  $(J-J_S^*)$. 
In addition we find a new manifestation of the universal behavior in the 
quantity $J_S^*$. According to the simple theory which leads to Eq.(\ref{p_s})
for grains with even (odd) $N$ this quantity is related to the renormalized 
two (three) levels pairing constant $\tilde{\lambda}_2$ ($\tilde{\lambda}_3$).
By fitting the linear part of the curves in Fig.\ref{peven} and 
Fig.\ref{podd}, we find that $\tilde{\lambda}_2$ and $\tilde{\lambda}_3$ 
are given by a universal functions of $\delta/\Delta$, of the form 
$\tilde{\lambda}_{2S}=\delta/\ln(a_S \delta/\Delta)$~\cite{matlar,mastellone}.
This result, shown in the insets of the Fig.\ref{peven} and Fig.\ref{podd},
is valid over several decades of values of the parameters.

\begin{figure}
\includegraphics[width=100mm,height=70mm]{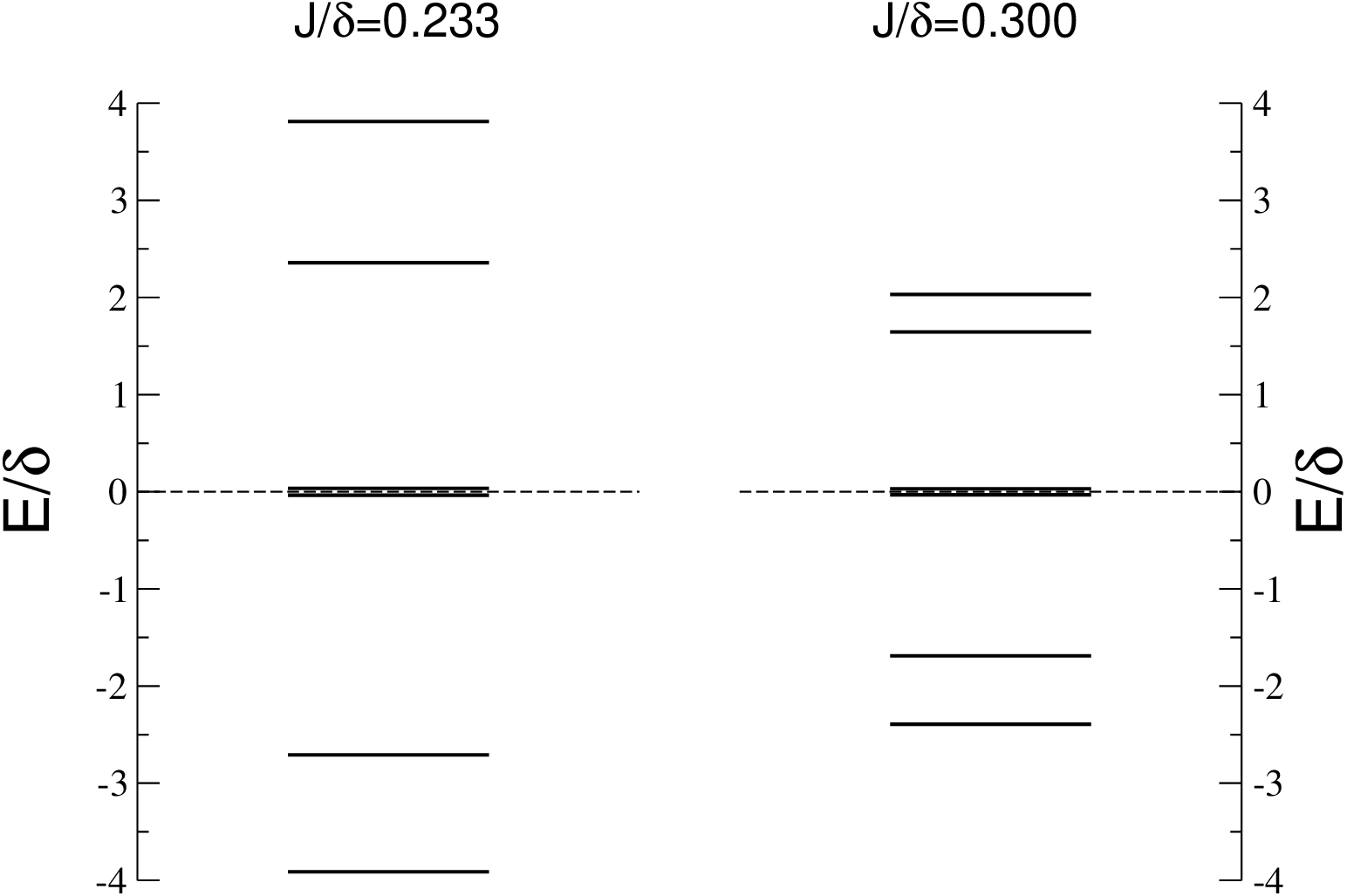}
\caption{\label{levconf} 
A typical configuration of the single-particle energy
levels (we chose as an example $\Omega=6$ and $\delta/\Delta=25$) which 
leads to a ground state with spin one in the tail region
(left side, $J/\delta=0.233$) and in the linear region (right side,
$J/\delta=0.300$). The Fermi level has been set to zero.}
\end{figure}

We now discuss more carefully the behavior near $J_1^*$ for grains with 
even $N$ (Fig.\ref{peven}). For samples with 
larger pairing interaction (smaller $\delta / \Delta$) the 
probability $P_1$ shows a non vanishing tail for $J \lesssim J_1^*$. 
We argue that this effect is due to sample to sample fluctuations of 
the {\em renormalized} $\tilde{\lambda}_{2S}$ which we ignored in 
Eq.(\ref{p_s}).
To understand this point consider even $N$ and samples for which the 
two levels in the central cluster at the Fermi energy are closely spaced 
($s_2 \ll \delta$). These samples may contribute to $P_1(J/\delta)$ for 
$J \sim J_1^*$. The two-level renormalized coupling in one of these grains is 
determined by the configuration of the other levels in the $\Omega$ shell 
and in particular depends strongly  
on the spacing $s_4$ between next neighboring pair of levels 
(above and below the Fermi energy, see the right side of Fig.\ref{levconf}). 
For most of the samples $s_4 \approx 3 \delta$ and they will have 
approximately the same two-level renormalized coupling, 
which is identified as the threshold 
$J_1^*$ in Eq.(\ref{p_s}). However for a small fraction of 
samples  we may have $s_4 \gg 3 \delta$ (see the left side of 
Fig.\ref{levconf}) which leads to a smaller value of the renormalized 
coupling. As a consequence
pairing correlations will be weaker and the ground state will have $S=1$ 
for a value of $J$ smaller than $J_1^*$. These samples determine
the appearance of the tail for $J \lesssim J_1^*$.

\begin{figure}
\includegraphics[width=100mm]{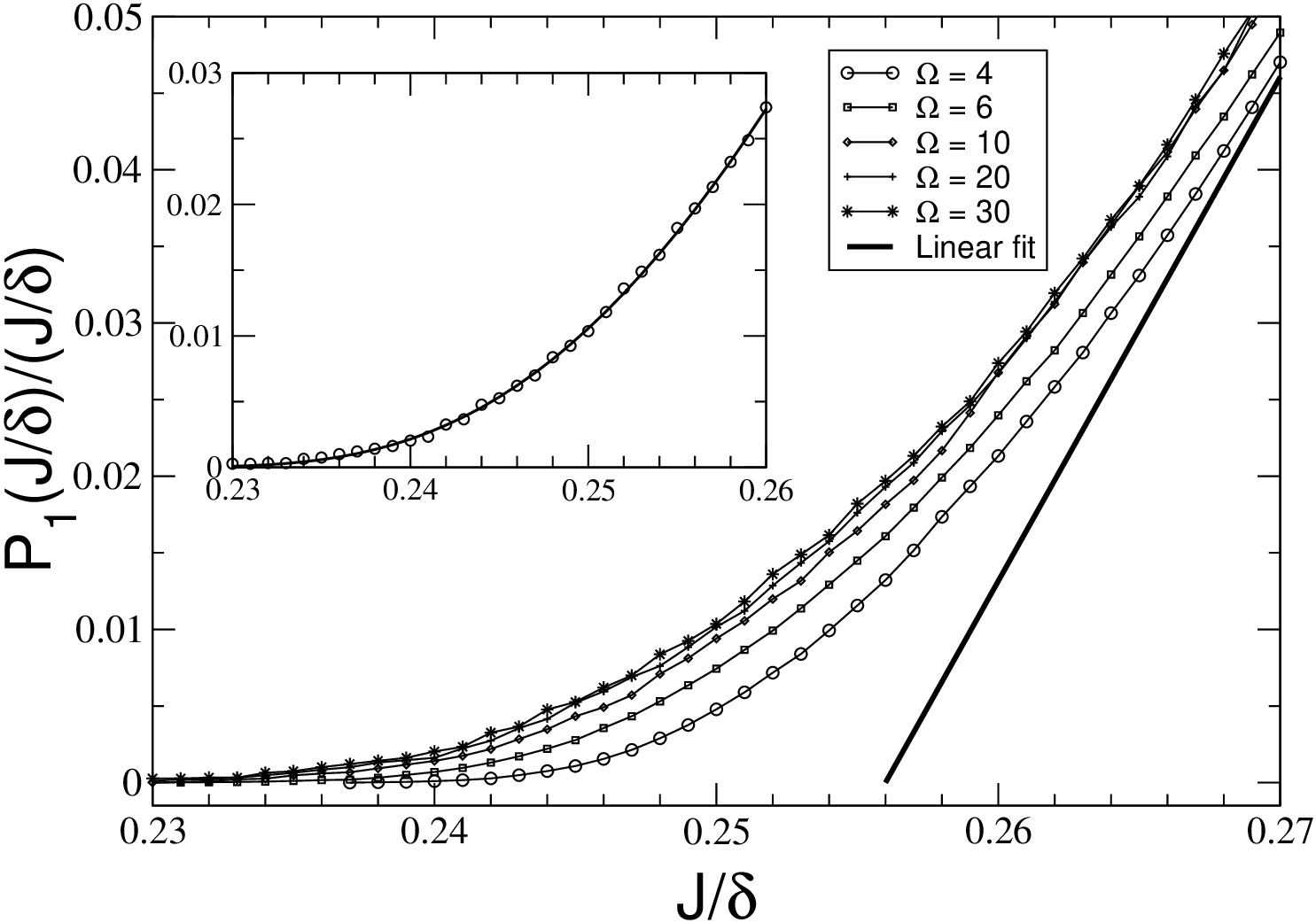}
\caption{\label{dD25} The behavior of the spin-1 ground
state probabilities for different values of the bandwidth
$\Omega$ (curves with different symbols) at the fixed value 
$\delta/\Delta=25$ in the tail region. The form 
$ P_1(J/\delta) =(81/64\pi^2\delta^2) J(J-\lambda)\mbox{erfc}[b/(J-\lambda)]$
($\mbox{erfc}(x)$ is the complementary error function) was chosen from the 
solution with three levels.
The fit to the curve with $\Omega=30$ (circles, in the inset) is obtained 
for $\lambda/\delta \sim \tilde{\lambda}_3/\delta=0.201$ and 
$b=6.821 \cdot 10^{-2}\delta$.}
\end{figure}

To check this argument we have first analyzed the typical level 
configurations of samples contributing to $P_1(J/\delta)$ (Fig.\ref{levconf}).
Indeed samples which contribute to $P_1(J/\delta)$ in the tail region have 
level configurations as in Fig.\ref{levconf}, with a 
central cluster of two levels around the Fermi energy, and
all the others very far away, whereas in the region described by 
Eq.(\ref{p_s}) the other levels are closer to
the two central ones. Further insight can be gained by including 
in Eq.(\ref{p_s}) the effect of fluctuations of a neighboring level.
We consider a shell of three levels with two electrons. Two levels are close
while the third lies far away.
In this system we determine the approximate form of the  
probability distribution $P_1(J/\delta)$ 
(caption of  Fig.\ref{dD25}), which is
non vanishing in the tail region.
Moreover we fitted this result with numerical data for larger systems (up to 
$\Omega=30$, see inset of Fig.\ref{dD25}), using 
the value of the renormalized coupling as a 
fitting parameter(see the inset of the Fig.\ref{dD25}). 
The agreement with the numerical data suggests that fluctuations of the 
renormalized pairing constant due to the statistics of far levels are 
negligible.

We also studied the behavior of $P_1(J/\delta)$ for $J \lesssim J^*_1$, for 
fixed $\delta/\Delta=25$ and 
for various values of the bandwidth $\Omega$. 
We used sets of GOE levels with $10^7$ realizations for 
$\Omega=4,6,10$ and with $10^6$ realizations for 
$\Omega=20,30$. We see that for the values of $J$ we could study 
$P_1(J/\delta)$
becomes independent on $\Omega$ for  $\Omega \gtrsim 20$, showing that 
still $P_1(J/\delta)$ depends on features of the universal level statistics. 
However by decreasing $J$ the behavior of $P_1(J/\delta)$ will depend more and 
more on details of the full $\Omega$ shell so universality is lost
(for instance $P_1(J/\delta)$ is expected to be nonzero for $J > \lambda$, 
the {\em bare} coupling).
On the contrary by increasing $J$ we recover the behaviour of
Eq.(\ref{p_s}) and all the curves collapse on the same line. 
The above analysis could be in principle carried out 
also in samples with odd $N$. However the unpaired electron in 
the spin 1/2 ground state weakens 
pairing correlations. As a consequence the tail in $ P_{3/2}(J/\delta)$ 
is tiny and it would require a much larger statistics to be investigated. 

\begin{figure}
\centering 
\includegraphics[width=100mm]{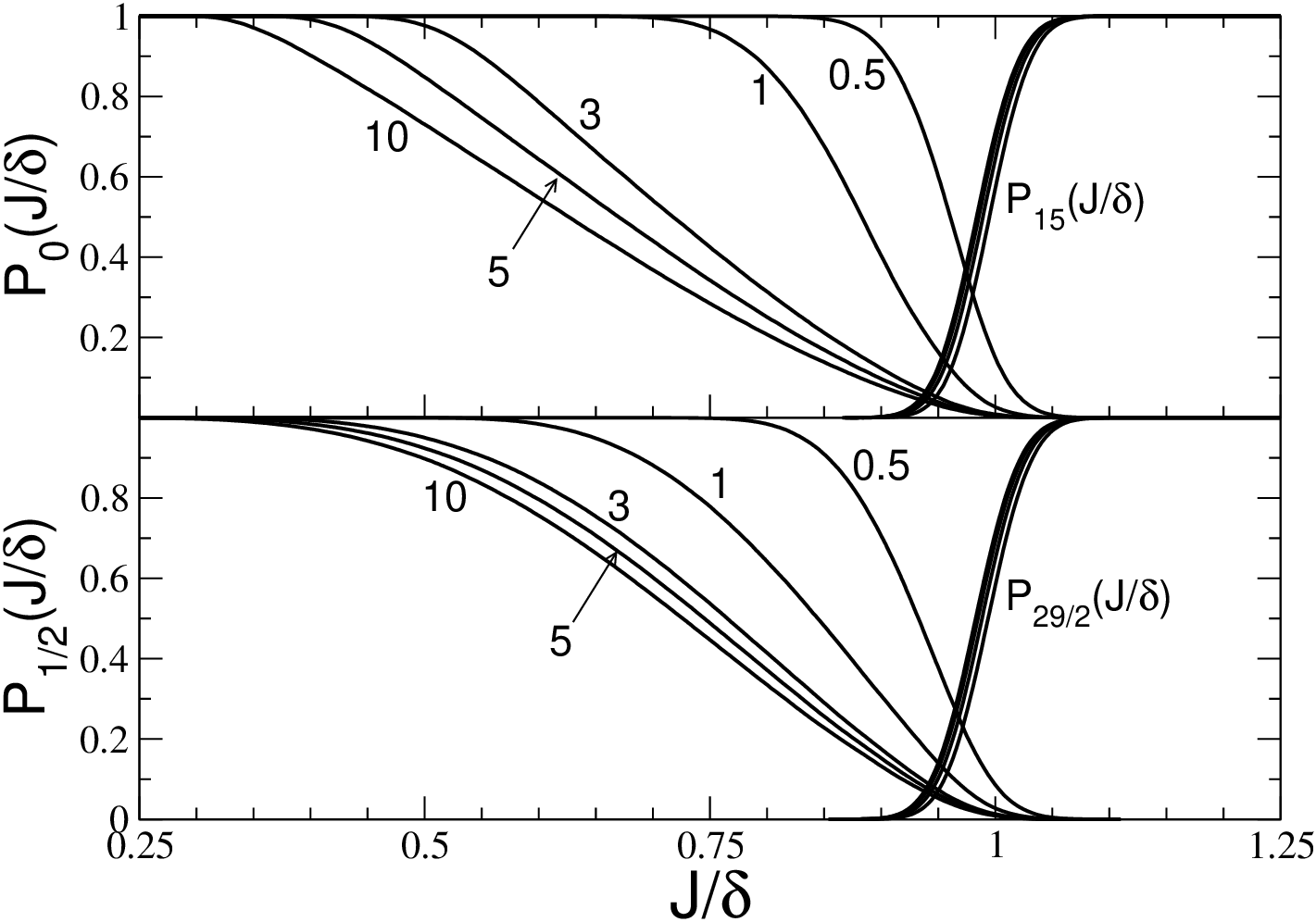} 
\caption{\label{trackevenodd} 
The behavior of the probability distributions in the systems
with an even (upper graph, $\Omega=30$) and odd (lower graph, $\Omega=29$) 
number of electron, of the smallest ($S=0$ and $S=1/2$, decreasing curves 
on the left side) and largest ($S=15$ and $S=29/2$, curve on right side
of the figure) spin ground states for different values of the $\delta/\Delta$ 
ratio (denoted by the numbers near each curve, in the case of largest spin 
value all the curves collapse on a single one) in a set of $10^5$ 
realizations.}
\end{figure}

Finally we consider grains with larger pairing interaction such that 
$\delta \lesssim \Delta$. This is the crossover region to BCS-like 
superconductivity~\cite{mastellone,berhal} and an  analytical approach 
to the problem is a formidable task, so we studied this regime numerically. 
Some qualitative results can be inferred by looking at the evolution of 
the behavior of the probability distributions $P_S(J/\delta)$ from the 
perturbative region $\delta \gg \Delta$ to the region $\delta \sim \Delta$.
In Fig.\ref{trackevenodd} the probability
distributions of the smallest and largest spin ground states 
are shown for the systems with an even ($\Omega=30$) and odd ($\Omega=29$) 
number of electrons in $10^5$ GOE realizations.
For smaller $J$ we have $P_0(J/\delta)=1$ 
whereas for large $J$ the maximal spin is 
favoured. In the intermediate region the stable value of $S$ is finite but the 
physics is sensitive to mesoscopic fluctuations. Results show that this 
crossover region becomes narrower as the ratio $\delta/\Delta$ decreases.
This trend is confirmed by the fact that, on increasing
the exchange, the system tends to maximize the spin in the ground state,
whereas in the dot regime $\delta > \Delta$ the crossover occurs gradually 
across all the increasing values of the spin. In the samples with an odd 
number of electrons the width of the $J$ range between the probability 
of the smallest and largest spin probability is larger than in the even 
systems since the pairing is weakened
by the presence of the unpaired electron. 

In conclusion we studied a model for the competition between superconducting 
and ferromagnetic ordering in small metal grains. The presence of pairing 
implies that all the levels in the shell are responsible for the determination
of the spin in the ground state. As compared to the normal case there are 
three new features: i) a soft threshold appears in the $P_S$ directly 
related to 
the renormalized pairing coupling; 
ii) the power law behavior has a new exponent as 
compared to the case where pairing is absent;
iii) at low enough exchange couplings $P_S(J/\delta)$ is determined by 
sample to sample fluctuations in the pairing coupling. By increasing both 
the pairing interaction 
and the exchange coupling we found that the region of the phase diagram 
where mesoscopic effects are important becomes progressively narrower.

\begin{acknowledgments}
We acknowledge very useful
discussions with B.L. Altshuler, L. Amico, A. Di Lorenzo, 
A. Fubini, and A. Osterloh. We acknowledge financial support from
European Community  (grant RTN2-2001-00440).
\end{acknowledgments}

\end{document}